\documentclass[aps,pra,twocolumn]{revtex4}

\usepackage{graphicx}
\usepackage{amsmath}

\usepackage{bm}
\usepackage{hyperref}
\usepackage{array}

\newcommand{\p}{\partial}

\newcommand{\ez}{\bm{\hat{e}}_3}

\newcommand{\emn}{\epsilon_{\mu\nu}}

\newcommand{\eln}{\epsilon_{\lambda\nu}}
\newcommand{\half}{\frac{1}{2}}

\newcommand{\lex}{\ell_{\rm ex}}

\newcommand{\Wa}{W_{\rm a}}
\newcommand{\hext}{h_{\rm ext}}
\newcommand{\bhext}{\bm{h}_{\rm ext}}
\newcommand{\Je}{J_{\rm e}}
\newcommand{\Jp}{J_{\rm p}}

\newcommand{\skyrmion}{n}
\newcommand{\Skyrmion}{\mathcal{N}}

\newcommand{\storque}{\beta}
\newcommand{\polarity}{\lambda}
\newcommand{\winding}{\kappa}
\newcommand{\VAsize}{d}

\newcommand{\bz}{{\bar{z}}}

\newcommand{\Omegax}{X}
\newcommand{\bOmegax}{\overline{X}}

\begin{document}

\title{Frequency generation by a magnetic vortex-antivortex dipole in spin-polarized current}

\author{Stavros Komineas}
\affiliation{Department of Applied Mathematics, University of Crete, 71409 Heraklion, Crete, Greece}

\begin{abstract}
A vortex-antivortex (VA) dipole may be generated due to a spin-polarized current flowing through a nano-aperture in a magnetic element. 
We study the vortex dipole dynamics using the Landau-Lifshitz equation in the presence of an in-plane applied magnetic field and a Slonczewski spin-torque term with in-plane polarization.
We establish that the vortex dipole is set in steady state rotational motion.
The frequency of rotation is due to two independent forces:
the interaction between the two vortices and
the external magnetic field.
The nonzero skyrmion number of the dipole is responsible for both forces giving rise
to rotational dynamics.
The spin-torque acts to stabilize the vortex dipole motion at a definite vortex-antivortex separation distance.
We give analytical and numerical results for the angular frequency of rotation 
and VA dipole features as functions of the parameters.
\end{abstract}


\maketitle

\section{Introduction}

The dynamics of the micromagnetic structure in ferromagnetic elements presents unusual features \cite{stoehr,malozemoff}.
Switching between different magnetic states which is a fundamental issue for magnetic recording, and magnetic oscillations which are exploited for the design of frequency generators \cite{Russek_Rippard_Cecil_2010} are examples for areas of applications where magnetization dynamics is of primary importance.
The most promising method to probe the dynamics of magnetic structures is the injection of spin-polarized current in magnetic nanoelements.

Of particular interest, in this respect, are
magnetic vortices which have been studied in the last years in thin magnetic nanoelements
\cite{kamenetskii08,Antos_Otani_Shibata_2008}.
The dynamical behavior of vortices has been linked to their topology 
\cite{malozemoff,huber82,nikiforov83,pokrovskii85,papanicolaou91,komineas96}.
The issue of vortex dynamics has been dramatically raised anew by recent experimental  observations.
In Ref.~\cite{Finocchio_Ozatay_PRB2008} a spin-polarized current injected into a magnetic element generated a microwave signal, attributed to a spontaneously nucleated VA pair in rotational motion \cite{komineas07}.
Spin-tranfer-driven oscillations due to an apparently nonuniform magnetic state were measured in \cite{rippard_pufall_kaka_PRL2004,pufall_rippard_schneider_PRB2007}. 
Synchronized motion of vortices could make the system a nanosized microwave generator \cite{ruotolo09}.
Vortex states have been proposed in connection to experimental data by micromagnetic simulations\cite{Berkov_Gorn_JAP2006,Berkov_Gorn_PRB2009}.
In another set of experiments switching of the vortex polarity was obtained
\cite{waeyenberge06,yamada07,Kammerer_Weigand_Schutz_2011,Pigeau_DeLoubens_Molenkamp_2011}, via dynamic generation of a vortex-antivortex (VA) pair \cite{hertel_gliga_2007,lee_guslienko_2007,kravchuk07}.

We will show by analysis and computations that a VA pair where the vortex and the antivortex have opposite polarities, under the influence of a spin-polarized current, is set in a stable rotational motion due to two independent forces: the internal interaction forces and an in-plane external field. The motion is stabilized by the spin-torque with in-plane spin-polarization.
We interpret the obtained frequency values through a virial relation.
The detailed theoretical study of the dynamics of nontrivial magnetization configurations is crucial for designing the current intensive experimental research.
We further expect that understanding frequency generation using robust configurations, like magnetic vortices, may lead to the design of devices.

The outline of the paper is as follows.
In Sec.~\ref{sec:llgs} we introduce the Landau-Lifshitz-Gilbert-Slonczewski (LLGS) equation.
In Sec.~\ref{sec:isotropic} we give a rotating VA pair solution of the pure exchange model.
In Sec.~\ref{sec:virial} we give a virial relation which contains the angular frequency of rotation.
In Sec.~\ref{sec:aperture} we give numerical results for rotating VA pairs when spin-polarized current is
flowing through a nano-aperture.
In Sec.~\ref{sec:conclusions} contains our concluding remarks.
In the Appendix we give numerical results used in the figures.

\section{The Landau-Lifshitz-Gilbert-Slonczewski equation}
\label{sec:llgs}

Our main tool is the Landau-Lifshitz-Gilbert-Slonczewski (LLGS) equation \cite{Berkov_Miltat_JMMM2007} which describes the dynamics of the magnetization $\bm{m}=(m_1,m_2,m_3)$ in ferromagnets including the effects of energy dissipation and spin-polarized current \cite{slonczewski96,berger96}.
The rationalized form of the equation reads
\begin{align}  \label{eq:llgs}
\dot{\bm{m}} & = 
 -\bm{m}\times ( \alpha_1\bm{f} -  \alpha_2\storque\,\bm{p} ) 
 -\bm{m}\times \left[\bm{m}\times (\alpha_2\, \bm{f} + \alpha_1\storque\,\bm{p}) \right],  \nonumber \\
 \bm{f} & = \Delta\bm{m} - m_3\ez + \bhext.
\end{align}
The dissipation constant is $\alpha$ and
$\alpha_1= 1/(1+\alpha^2),\; \alpha_2 =\alpha\alpha_1$,
the first term in $\bf{f}$ is the exchange field, in the second (anisotropy) term $\ez$ is the unit vector along the third magnetization direction so it describes easy-plane anisotropy in the $(x,y)$ plane which is considered to be the plane of the film, and $\bhext$ is an external magnetic field.
The easy-plane anisotropy is considered mainly as a simplified model for the magnetostatic interaction.
This is an approximation for thin films and it cannot be a substitute for full micromagnetic simulations.
The magnetization $\bm{m}$ and the field $\bhext$ are measured in units of the saturation magnetization $M_s$, so we have $\bm{m}^2=1$.
We consider a simplified form of the spin-torque term where $\storque$ is a constant, and $\bm{p}$ is the unit vector in the polarization direction of the current.
We have $\storque = \Je/\Jp,\;\Jp = \mu_0 M_s^2 |e| d_f/\hbar$ where $d_f$ is the film thickness and $\Je$ is the current density.
As an example, for a permalloy film with $d_f=5\,{\rm nm}$ we have $\Jp=4.54\times e^{-12}\,{\rm A/m}^2$.
The unit of length (exchange length) in Eq.~\eqref{eq:llgs} is $\lex = \sqrt{2A/(\mu_0 M_s^2)}$
where $A$ is the exchange constant,
and the unit of time is $\tau_0 = 1/(\gamma \mu_0 M_s)$
where $\gamma$ is the gyromagnetic ratio.
For permalloy we have
$\lex = 7.00\,{\rm nm},\; \tau_0 = 6.96\,{\rm psec}$ implying a frequency $f_0=22.9\,{\rm GHz}$.

Eq.~\eqref{eq:llgs} has static vortex solutions for $\storque=0,\,\bhext=0$.
A vortex is a magnetization configuration characterized by a {\it winding number} $\winding=1$ while an antivortex has $\kappa=-1$.
A further topological invariant is the {\it skyrmion number} which is defined as the integral over the whole plane
\begin{equation}  \label{eq:skyrmion}
\Skyrmion = \frac{1}{4\pi} \int \skyrmion\,d^2x,\qquad \skyrmion = \half \emn (\p_\nu \bm{m} \times \p_\mu \bm{m})\cdot \bm{m},
\end{equation}
where  $\mu,\nu=1,2$, $\emn$ is the totally antisymmetric tensor, and $\skyrmion$ is called the local vorticity \cite{papanicolaou91}.
For a vortex the skyrmion number is $\Skyrmion=-\kappa\polarity/2$ where $\polarity=\pm 1$ is the vortex polarity.
Consequently, a VA pair where the vortex has negative polarity and the antivortex positive polarity has $\Skyrmion=1$.
Despite their nonzero $\Skyrmion$ such VA pairs can be generated by finite energy processes,
such as that numerically observed in Ref.~\cite{Berkov_Gorn_PRB2009} in Fig.~13c.

We will consider a thin film (or a sufficiently large magnetic element) with boundary condition
\begin{equation}  \label{eq:bc}
\bm{m}(|\bm{r}| \to\infty) = (1,0,0)
\end{equation}
consistent with a VA pair configuration.
Such a boundary condition may be due to the shape of the film 
or it may be imposed by the application of an external magnetic field of the form
\begin{equation}  \label{eq:hext}
\bhext = (\hext,0,0).
\end{equation}
For the spin-current term we use
\begin{equation}  \label{eq:p}
\storque\, \bm{p} = \storque\,  (1,0,0),\qquad  \storque < 0,
\end{equation}
which favours $\bm{m}=(-1,0,0)$ opposite to the magnetization at the boundary.
Such polarization was used in Ref.~\cite{Finocchio_Ozatay_PRB2008} at a central region of the element
and generation of a VA pair of opposite polarities was observed.

\section{The isotropic model}
\label{sec:isotropic}

We will first investigate Eq.~\eqref{eq:llgs} for zero anisotropy, that is we set $\bm{f}=\Delta\bm{m}+\bhext$.
A convenient formulation of the problem is obtained using the complex variable
\begin{equation}  \label{eq:variableX}
\Omegax = \frac{m_2+i m_3}{1-m_1}
\end{equation}
where the real and imaginary parts give the stereographic projection of $\bm{m}$ from the point $\bm{m}=(1,0,0)$.
We consider $\Omegax=\Omegax(z,\bz)$ where $z=x+iy$ is the complex position on the $(x,y)$ plane of the film, and $\bz$ is its complex conjugate. The equation of motion reads
\begin{equation}  \label{eq:exchange_model}
(i-\alpha)\,\dot{\Omegax} = -4\,\p_z \p_{\bz} \Omegax 
 + \frac{8\bOmegax}{1+\Omegax\bOmegax}\,\p_z \Omegax\, \p_{\bz} \Omegax
- (\hext - i\storque)\,\Omegax.
\end{equation}

For $\hext=0,\;\storque=0$
any function $\Omegax=\Omegax(z)$ is a static solution \cite{belavin75}.
The simple solution
$\Omegax = z/a_0$,
where $a_0$ is a complex constant, respects the boundary condition \eqref{eq:bc}. It represents a vortex with negative polarity centered at $-ia_0$ (where $m_3=-1$) and an antivortex with positive polarity centered at $ia_0$ (where $m_3=1$), and we have $\Skyrmion=1$. The constant $|a_0|$ gives vortex core size while the distance between the vortex and the antivortex is $\VAsize=2 |a_0|$ \cite{gross78}.

Let us now consider nonzero and uniform $\hext$ and $\storque$. For $\Omegax(t=0)=z/a_0$ Eq.~\eqref{eq:exchange_model} has the solution
\begin{align}  \label{eq:VAx_time}
& \Omegax(z,t) = \frac{z}{a(t)},  \\
& a(t) = a_0\, \exp\!{\Big[ -\alpha_1 \left[(\alpha \hext + \storque) + i (\hext - \alpha\storque)\right]\,t \Big]}. \nonumber
\end{align}
The distance between vortex and antivortex is 
$\VAsize(t) = 2 |a(t)| = 2 |a_0|\, \exp\!{\left[ \alpha_1 (\alpha \hext + \storque)\,t \right]}$,
while the vortices rotate around the origin at an angular frequency
$\omega = -\alpha_1\,(\hext-\alpha\storque)$.
The condition $\alpha\hext+\storque=0$ gives balance of radial forces for, e.g., $\hext > 0$ and $\storque < 0$.
For this interesting special case we have a rotating VA pair in steady state with an angular frequency
$\omega=-\hext$. For the case $\Omegax(t=0)=\bz/a_0$ which has $\Skyrmion=-1$ 
we obtain rotation in the opposite sense.
So we conclude that the angular frequency is
\begin{equation}  \label{eq:omega_BP}
\omega=-\hext/\Skyrmion,
\end{equation}
for $\Skyrmion=\pm 1$.
We have thus established that a VA pair is set in steady state rotational motion due to external field and spin-torque.
Notably, the in-plane field  $\bhext$, typically expected to induce magnetization precession around its direction, is actually giving rotation of a magnetization configuration with $\Skyrmion=1$ around the axis perpendicular to the film.

\section{A virial relation}
\label{sec:virial}

Let us now return to the full Eq.~\eqref{eq:llgs} and assume the existence of a rigidly rotating VA pair in steady state.
This is expressed by the relation
\begin{equation}  \label{eq:rigid_rotation}
\dot{\bm{m}} = -\omega\, \eln\,x_\lambda \p_\nu\bm{m},
\end{equation}
where $\lambda, \nu=1,2$ and $\eln$ is the totally antisymmetric tensor.
Eq.~\eqref{eq:rigid_rotation} is inserted in Eq.~ \eqref{eq:llgs} which gives virial (integral) relations.
For a uniform magnetic field \eqref{eq:hext} and spin-torque polarization \eqref{eq:p}
a, so-called, Derrick relation is obtained \cite{inpreparation}
\begin{align}  \label{eq:derrick}
  \omega & \left( \ell + \frac{\alpha}{2} \int \epsilon_{\lambda\nu}\, x_\lambda x_\mu d_{\mu\nu}\,d^2x \right) = \nonumber \\
     & - \left( \Wa + \hext\,\mu_1 + \half \int x_\mu \tau_\mu\,d^2x \right),
\end{align}
where
$\Wa = \frac{1}{2} \int (m_3)^2\,d^2x$ is the anisotropy energy,
$d_{\mu\nu} = \p_\mu \bm{m}\cdot \p_\nu \bm{m},\;
\tau_\mu = -\storque(\bm{m}\times\p_\mu\bm{m})\cdot\bm{p}$,
and the integrals extend over the whole plane.
The quantity
\begin{equation}  \label{eq:angular_momentum}
 \ell = \half \int \rho^2\,\skyrmion\,d^2x,\qquad \rho^2 = x_1^2+x_2^2,
\end{equation}
gives a measure of the size of the VA pair and is identified with the angular momentum \cite{papanicolaou91}.
We have also defined
\begin{equation}  \label{eq:magnetization1}
 \mu_1= -\frac{1}{2} \int x_\mu \p_\mu m_1\, d^2x = \int (1-m_1)\,d^2x,
\end{equation}
where the last equation derives from a partial integration assuming vanishing boundary terms,
and the last quantity gives the total magnetization (spin reversals) in the negative $x$ direction.

Let us verify Eq.~\eqref{eq:derrick}
in the case of the exchange model, where we should set $\Wa=0$.
For the rotating solution \eqref{eq:VAx_time} we have $\int x_\mu \tau_\mu\,d^2x=0$ and $d_{12}\!=\!d_{21}\!=\!0,\; d_{11}\!=\!d_{22}$ thus $\epsilon_{\lambda\nu}\, x_\lambda x_\mu d_{\mu\nu}=0$.
The Derrick relation is now greatly simplified and gives the angular frequency as
\begin{equation}  \label{eq:derrick_exchange}
   \omega = -\hext\,\frac{\mu_1}{\ell}.
\end{equation}
We have
\begin{equation}  \label{eq:ell=mu1}
 \ell = \half \int \frac{4 a^2\, \rho^2}{(\rho^2+a^2)^2}\, (2\pi\rho d\rho) = \mu_1,
\end{equation}
where the first form for $\mu_1$ in Eq.~\eqref{eq:magnetization1} is used.
The integral in Eq.~\eqref{eq:ell=mu1} diverges if it extends over the whole plane. However, the Derrick relation is valid also when the integrations in Eq.~ \eqref{eq:derrick} extend over finite regions, except some boundary integrals are added which go to zero for large integration regions.
We finally conclude that Eq.~\eqref{eq:derrick_exchange} gives $\omega=-\hext$ for the solution \eqref{eq:VAx_time}, as expected.

It is useful to note that the form \eqref{eq:derrick_exchange}  can be obtained for configurations which have the following symmetries: $m_1$ is even in $x$ and $y$, $m_2$ is odd in $y$ and even in $x$, and $m_3$ is odd in $x$ and even in $y$ (also, when $m_2, m_3$ exchange symmetries).
We call VA pair configurations with such symmetries {\it symmetric VA pairs}.

\begin{figure}[t]
   \centering
   \includegraphics[width=3.1in]{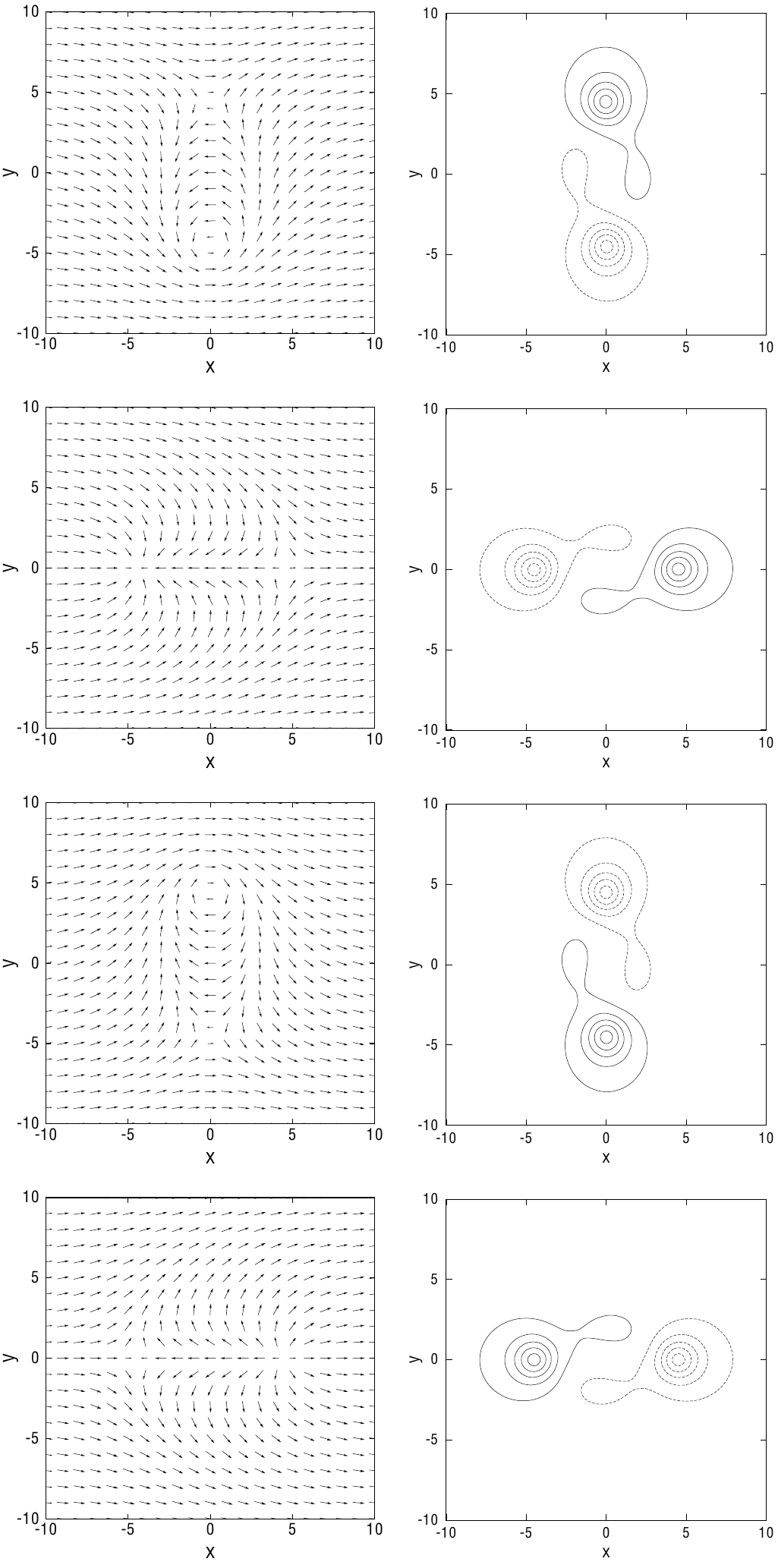}
   \caption{VA pair in steady state rotation for spin-polarized current through an aperture with parameter set \eqref{eq:parameters2} and applied field $\hext=0.05$.
   We show snapshots at times (from top) $t=0, 23, 46, 69$, while the period of rotation is $T \simeq 91.5$.
Left column shows the vector $(m_1, m_2)$. 
Right column shows contour plots for $m_3=\pm 0.1, \pm 0.3, \pm 0.5, \pm 0.7, \pm 0.9$ with positive values 
represented by solid lines and negative values by dashed lines. We have an apparent precession of the 
magnetization in the vicinity of the vortex and the antivortex as they rotate (clockwise).}
   \label{fig:config_a002b00200h0050}
\end{figure}

\section{Spin-current through a nano-aperture}
\label{sec:aperture}

We proceed to apply the ideas developed so far to an experimental setup where spin-polarized current is injected in a nanoelement through a nano-aperture \cite{Finocchio_Ozatay_PRB2008}.
Numerical simulations have shown that a VA pair of opposite polarities is spontaneously created and its rotating motion generates microwave frequencies.
The permalloy nanoelements of thickness $5\,{\rm nm}$ and have an elliptic shape. The shape apparently favors alignment of the magnetization along the major axis of the ellipse effectively imposing, to a rough approximation, boundary condition similar to \eqref{eq:bc}.
The external field used is in the range $|H_{\rm ext}| < 25\,{\rm mT}$ which, in the units of Eq.~\eqref{eq:llgs}, reads $|\hext| < 0.03$.
(since $\mu_0 M_s = 0.817\,{\rm T}$).
A current $I=-4\,{\rm mA}$ through an aperture of diameter $d_a=40\,{\rm nm}$ corresponds to 
a current density $\Je=3.18 \times 10^{-12}\,{\rm A/m}^2$ and 
$\storque=-0.70$ (but partial polarization of the current should effectively give a smaller $|\storque|$).

We have simulated numerically Eq.~\eqref{eq:llgs} using an external field \eqref{eq:hext} and a spin-torque term \eqref{eq:p} which is nonzero only in the circular region of an aperture with diameter $d_a$. We will present results for the parameter values
\begin{equation}  \label{eq:parameters2}
\storque=-0.20,\qquad \alpha=0.02,\qquad d_a=6\,\lex.
\end{equation}
An initial VA pair configuration evolves under Eq.~\eqref{eq:llgs} and it relaxes to a 
steady state rotating VA pair for a range of values of the external field $\hext$.
Fig.~\ref{fig:config_a002b00200h0050} shows snapshots of a rotating VA pair in steady state for the parameter set \eqref{eq:parameters2} and $\hext=0.05$.
We have an apparent precession of the magnetization in the vicinity of the vortex and the antivortex.
It is important to realize that this process is absolutely smooth, and thus perfectly realizable.

This situation should be contrasted to
rotating VA pairs in the conservative Landau-Lifshitz (LL) equation ($\alpha,\storque,\hext=0$) which are apparently unstable states.
In the presence of dissipation they shrink until they degenerate to a point.
In summary, a VA pair in the conservative model rotates due to the interaction between the vortex and the antivortex, and the rotational dynamics is linked to the topology of the VA pair, i.e., its nonzero skyrmion number \cite{komineas07}.

\begin{figure}[t]
   \includegraphics[height=1.7in]{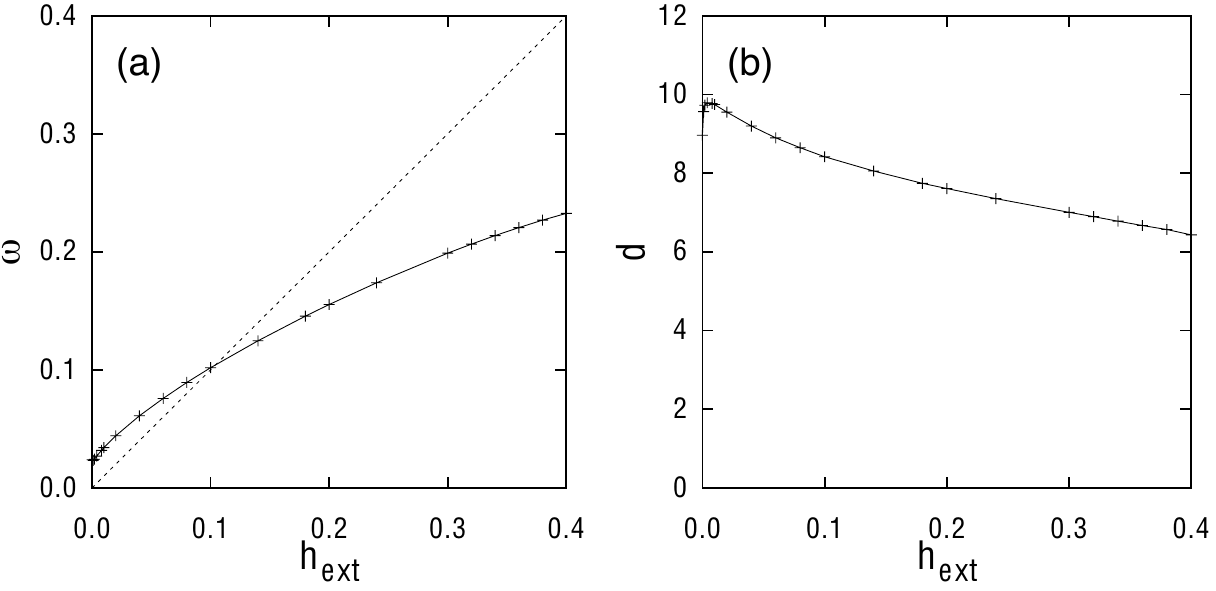}
   \caption{(a) Angular frequency of rotation $\omega$ as a function of the applied field $\hext$ for parameter set \eqref{eq:parameters2}. 
   Numerical results are shown by points connected by a line.
   The dotted line $\omega=\hext$ is plotted for comparison. Rotation is clockwise.
   (b) The vortex-antivortex separation distance $\VAsize$ as a function of the applied field $\hext$ for parameter set \eqref{eq:parameters2}.
Numerical results are shown by points connected by a line.
}
   \label{fig:hext_omega-position}
\end{figure}

\begin{figure}[t]
   \centering
   \includegraphics[height=1.7in]{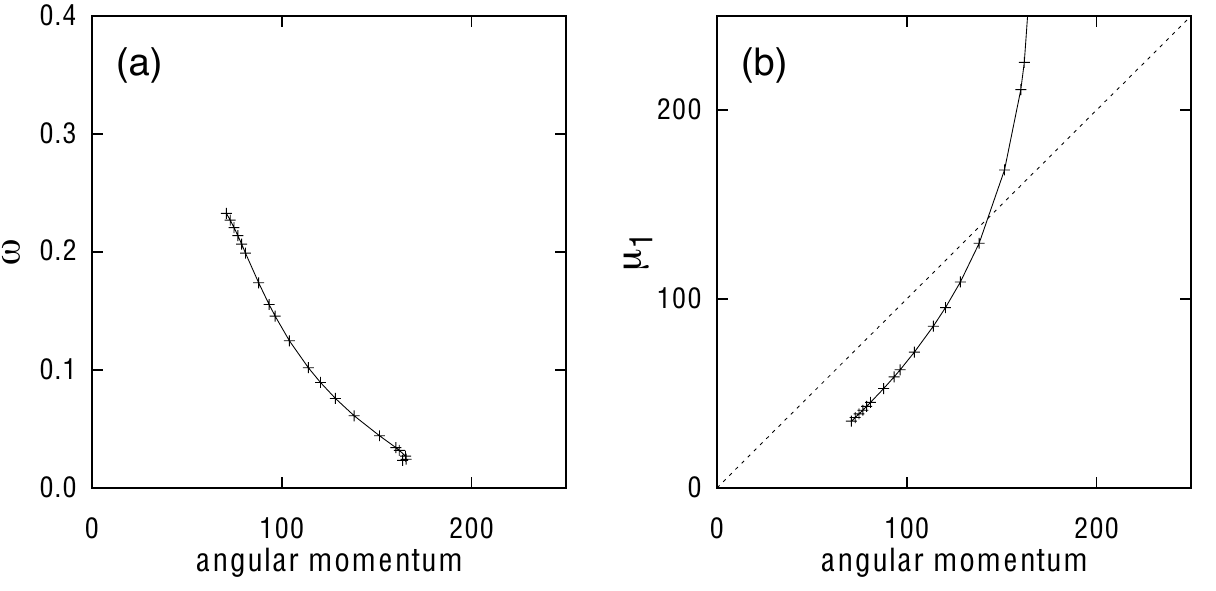}
   \caption{Angular frequency of rotation $\omega$ as a function of the VA pair angular momentum $\ell$, for parameter set \eqref{eq:parameters2}. Numerical results are shown by points connected by a line.
   The dotted line $\mu_1=\ell$ is plotted for comparison.
}
   \label{fig:ell_omega-mu}
\end{figure}

For the full LLGS equation, solution \eqref{eq:VAx_time} indicates that we have rotation of a VA pair due to in-plane external field and the spin-torque term.
This source of rotation is additional to the rotation due to interactions.
Fig.~\ref{fig:hext_omega-position}a shows the angular frequency of rotation
for $0 < \hext \leq 0.40$. Fig.~\ref{fig:hext_omega-position}b shows the VA separation distance $\VAsize$ as a function of $\hext$ and the dependence is seen to be weak.
As vortex positions we assume the points where $m_3=\pm 1$.
These are located outside the aperture for all values of $\hext$ simulated.
We do find a steady state also for $\hext=0$, and this has a particularly large $\mu_1$ compared to VA pairs for $\hext > 0$.
For $\hext > 0.40$ we do find rotating VA pairs but the radius of rotation fluctuates and a simple steady state is not reached \cite{Berkov_Gorn_PRB2009}.
We actually find fluctuations also for $\hext=0.38, 0.40$ in some simulations (depending on the initial condition), and this indicates hysteresis.

Fig.~\ref{fig:ell_omega-mu}a shows the angular frequency $\omega$ as a function of the angular momentum of the VA pair $\ell$.
For well separated vortices we expect \cite{komineas07}
\begin{equation}  \label{eq:ell_d}
\ell \approx \Skyrmion\frac{\pi}{2}\,d^2.
\end{equation}
We find that formula \eqref{eq:ell_d} gives indeed a good approximation to the data.

In order to derive a guide for the understanding of the angular frequency $\omega$ we return to the Derrick relation \eqref{eq:derrick}.
The latter is satisfied with an accuracy better than 1\% in all our numerical simulations.
The numerical data show that the term due to dissipaton (second term on the lhs) and the term containing $\storque$ (third term on the rhs) are negligibly small in all cases. This may be due to the relatively small asymmetry of the simulated VA pairs.
The Derrick relation gives for $\omega$ the approximate formula
\begin{equation}  \label{eq:derrick_approx}
  \omega \approx - \left( \frac{\Wa}{\ell} + \hext\,\frac{\mu_1}{\ell}  \right).
\end{equation}
For a free VA pair ($\hext, \storque=0$ and $\alpha=0$) we have $\omega=-\Wa/\ell$ \cite{komineas07}.
The anisotropy energy $\Wa$ for 
a static isolated vortex in the LL equation (assuming no external fields) 
is equal to $\pi/2$ \cite{komineas98}, so we expect $\Wa=\pi$ for a vortex pair.
Using the formula \eqref{eq:ell_d} for the angular momentum we obtain the approximate relation $\omega \approx -\Skyrmion (2/d^2)$ (for $\hext=0$).

For understanding the angular frequency generated by the vortex dipole
we schematically consider the two contributions $\omega=\delta\omega_1+\delta\omega_2$ in Eq.~\eqref{eq:derrick_approx}. We find $2.5 < \Wa < 3.5$ for the range of $\hext$ simulated, so we argue that the contribution $\delta\omega_1=\Wa/\ell$ is inversely proportional to $\ell$, as in the case of free VA pair rotation.
The contribution $\delta\omega_2$ of the second term on the rhs depends on
the relation of $\mu_1$ versus $\ell$. This is shown in Fig.~\ref{fig:ell_omega-mu}b, where we see that $\mu_1$ grows faster than linear with $\ell$. It suggests a contribution to $\omega$ due to the external field which is $\delta\omega_2 < \hext$ for $\ell<145$ while $\delta\omega_2 > \hext$ for $\ell>145$.
For a different interpretation of Fig.~\ref{fig:ell_omega-mu}a consider the following.
From the numerical data we see that $\delta\omega_2$ is the main contribution except for small values of $\hext$ (i.e., large $\ell$). One could thus take $\omega/\hext$ as an approximation for the ratio $\mu_1/\ell$.

We have performed a set of numerical simulations where we keep the applied field constant and vary the spin current strength. We briefly report here numerical results for
\begin{equation}  \label{eq:parameters3}
\hext=0.05\qquad \alpha=0.02,\qquad d_a=6.
\end{equation}
We find steady state rotating pairs for $|\storque| \geq 0.016$.
For decreasing values of $|\storque|$ the VA distance $\VAsize$ decreases and consequently the angular frequency increases.
We find a maximum $\omega=0.14$ and minimum VA distance $d=5$  (for the minimum $|\storque|$ value).
Note that the $\VAsize$ be smaller than $d_a$ in this case.

Finally, we comment on the rotational dynamics in the presence  of the magnetostatic field.
We discuss briefly how the picture in Fig.~\ref{fig:config_a002b00200h0050} would be modified.
An important point is that the vortices in the first and third entries (with radial orientation of the magnetization vector) would have higher magnetostatic energy.
It is straightforward to realize that the rotation of vortices would be accompanied by a rotation of the magnetization vector so that the vortex would tend to approach azimuthal magnetization orientation in its vicinity.
As the precession between radial and azimuthal vortex happens every half cycle, we would have the production of a frequency twice that of the rotating state discussed in this paper.
In the general case,  we expect that both frequencies $\omega$ and $2\omega$ would appear in the spectrum,
as measured in various experiments, e.g., in Refs.~\cite{rippard_pufall_kaka_PRL2004,Finocchio_Ozatay_PRB2008}.

In Ref.~\cite{Berkov_Gorn_PRB2009} rotating VA pairs in a full micromagnetic model are presented and they are accompanied by a satellite VA pair which is created and annihilated during a cycle of the main pair (the mode is termed $L_1$).
This process is directly compared to the comments of the previous paragraph.
Berkov et~al use numerical data to apply model \eqref{eq:derrick_approx} for $\hext=0$ and find that it predicts a frequency $\delta f_1=0.44\,{\rm GHz}$.
Eq.~\ref{eq:derrick_approx} indicates that a contribution approximately $\delta f_2=\hext\,f_0$ should be added to the previous number. Using the parameters of Ref.~\cite{Berkov_Gorn_PRB2009} we have $\hext=20\,{\rm mT}/(\mu_0 M_s)=0.00373$ and $f_0= \gamma \mu_0 M_s/(2\pi)=22.54\,{\rm GHz}$, which give $\delta f_2=84\,{\rm MHz}$.
The frequency is expected to be $f=2(\delta f_1+ \delta f_2)=1.05\,{\rm GHz}$ in good agreement with the reported value ($1.2\,{\rm GHz}$) from the simulation.

\section{Conclusions}
\label{sec:conclusions}

In conclusion, we have studied the dynamics of vortex-antivortex pairs of opposite polarities
using the Landau-Lifshitz-Gilbert-Slonczewski equation \cite{Berkov_Miltat_JMMM2007}.
The VA pair is set in rotational motion due to two independent forces: the internal interaction forces between vortex and antivortex and the external field, while the motion is stabilized by the spin-torque of polarized current.
Both the polarization of the current and the applied magnetic field are in-plane.
We have given analytical and numerical results for VA pairs which are in a stable steady-state rotation.
We believe that our results can be used as a framework for the description of 
frequency generation by topological solitons under probes, particularly, spin-polarized current.

A significant step forward would be obtained if 
one includes the magnetostatic field in Eq.~\eqref{eq:llgs} thus making the model more realistic.
We expect  rotational dynamics of a significantly more complex nature, but the gross feature would still be captured by the present framework.
The magnetostatic interaction decouples rotations in physical space by rotations in the magnetization space
and could thus reveal phenomena not manifested by the present model.

\section*{Aknowledgements}

This work was partially supported by the FP7-REGPOT-2009-1 project ``Archimedes Center for Modeling, Analysis and Computation'', and by grant KA3011 of the University of Crete.
I am grateful to Nikos Papanicolaou for many discussions and suggestions,
Giovanni Finocchio and Dima Berkov for discussions of numerical and experimental results and Luis Torres at the U. of Salamanca for hospitality and discussions. 

\appendix

\section{Numerical results}
\label{app:numerical_results}

We give in Table \ref{table:storque020_aperture} part of the numerical results used to plot Figs.~\ref{fig:hext_omega-position}, \ref{fig:ell_omega-mu}.
The angular frequency is calculated using the Derrick relation \eqref{eq:derrick} (it is in excellent agreement
with the frequency directly measured in the simulations). 

\begin{table}[h]
\begin{center}
\begin{tabular}{|c|c|c|c|c|}
\hline
\hspace{15pt}$\hext$\hspace{15pt}  & \hspace{15pt}$\VAsize$\hspace{15pt}  &  \hspace{15pt}$\ell$\hspace{15pt}   & \hspace{15pt}$\mu_1$\hspace{15pt}  & \hspace{15pt}$\omega$\hspace{15pt}   \\
\hline
 0.000  &    8.96  & 156.6 & 579.2  &  0.02378  \\
 0.004  &    9.79  & 165.2  &  272.4  &  0.02694  \\
 0.010  &    9.74  & 160.1  &  210.9  &  0.03409  \\
 0.040  &    9.20  & 138.2  &  129.6  &  0.06110  \\
 0.100  &    8.42  & 114.0   & 85.6  &  0.10175  \\
 0.200  &    7.61  &  93.3   &  58.8  &  0.15532  \\
 0.300  &    7.01  &  80.9   &  45.2  &  0.19884  \\
 0.400  &    6.42  &  70.9   &  35.3  &  0.23256  \\
\hline
\end{tabular}
\end{center}
\caption{Results of numerical simulations for spin current through an aperture and parameter set \eqref{eq:parameters2} ($\storque=-0.20, \alpha=0.02, d_a=6$).
We show the distance between the vortex and the antivortex $\VAsize$, the angular momentum $\ell$ and
the total magnetization $\mu_1$.
}
\label{table:storque020_aperture}
\end{table}


\begin{thebibliography}{10}

\bibitem{stoehr}
J. St\"ohr and H.~C. Siegmann, {\em Magnetism, From Fundamentals to Nanoscale
  Dynamics} (Springer, Berlin, 2006).

\bibitem{malozemoff}
A.~P. Malozemoff and J.~C. Slonczewski, {\em Magnetic Domain Walls in Bubble
  Materials} (Academic Press, New York, 1979).

\bibitem{Russek_Rippard_Cecil_2010}
S.~E. Russek, W.~H. Rippard, T. Cecil, and R. Heindl, {\em Spin-Transfer Nano-Oscillators}, in {\em Handbook of Nanophysics} (CRC Press, Honolulu, USA, 2010),
  Chap.~38, pp.\ 1--23.

\bibitem{kamenetskii08}
E. Kamenetskii~(ed), {\em Electromagnetic, Magnetostatic, and
  Exchange-Interaction Vortices in Confined Magnetic Structures} (Research
  Signpost, India, 2008).

\bibitem{Antos_Otani_Shibata_2008}
R. Antos, Y. Otani, and J. Shibata, J. Phys. Soc. Japan {\bf 77},  031004
  (2008).

\bibitem{huber82}
D.~L. Huber, Phys. Rev. B {\bf 26},  3758  (1982).

\bibitem{nikiforov83}
A. Nikiforov and E. Sonin, JETP {\bf 58},  373  (1983).

\bibitem{pokrovskii85}
V.~L. Pokrovskii and G.~V. Uimin, JETP Lett. {\bf 41},  128  (1985).

\bibitem{papanicolaou91}
N. Papanicolaou and T.~N. Tomaras, Nucl. Phys. B {\bf 360},  425  (1991).

\bibitem{komineas96}
S. Komineas and N. Papanicolaou, Physica D {\bf 99},  81  (1996).

\bibitem{Finocchio_Ozatay_PRB2008}
G. Finocchio, O. Ozatay, L. Torres, R. Buhrman, D. Ralph, B. Azzerboni, Phys. Rev. B {\bf 78},  174408  (2008).

\bibitem{komineas07}
S. Komineas, Phys. Rev. Lett. {\bf 99},  117202  (2007).

\bibitem{rippard_pufall_kaka_PRL2004}
W.~H. Rippard, M.~R. Pufall, S. Kaka, S.~E. Russek, T.~J. Silva, Phys. Rev. Lett. {\bf 92},  027201  (2004).

\bibitem{pufall_rippard_schneider_PRB2007}
M.~R. Pufall, W.~H. Rippard, M.~L. Schneider, and S.~E. Russek, Phys. Rev. B
  {\bf 75},  140404  (2007).

\bibitem{ruotolo09}
A. Ruotolo, V. Cros, B. Georges, A. Dussaux, J. Grollier, C. Deranlot, R. Guillemet, K. Bouzehouane, S. Fusil, A. Fert, Nature Nanotechnology {\bf 4},  528   (2009).

\bibitem{Berkov_Gorn_JAP2006}
D.~V. Berkov and N.~L. Gorn, Journal of Applied Physics {\bf 99},  08Q701
  (2006).

\bibitem{Berkov_Gorn_PRB2009}
D.~V. Berkov and N.~L. Gorn, Phys. Rev. B {\bf 80},  064409  (2009).

\bibitem{waeyenberge06}
B. Van Waeyenberge, A. Puzic, H. Stoll, K. W. Chou, T. Tyliszczak, R. Hertel, M. F\"ahnle, H. Br\"uckl, K. Rott, 
G. Reiss, I. Neudecker, D. Weiss, C. H. Back \& G. Sch\"utz, Nature(London) {\bf 444},  461  (2006).

\bibitem{yamada07}
K. Yamada, S. Kasai,5, Y. Nakatani, K. Kobayashi, H. Kohno, A. Thiaville \& T. Ono, Nature Materials {\bf 6},  269  (2007).

\bibitem{Kammerer_Weigand_Schutz_2011}
M. Kammerer, M. Weigand, M. Curcic, M. Noske, M. Sproll, A. Vansteenkiste,	 B. Van Waeyenberge,	 H. Stoll,	 G. Woltersdorf, C. H. Back	 \& G. Schuetz, Nat. Commun. {\bf 2},  279  (2011).

\bibitem{Pigeau_DeLoubens_Molenkamp_2011}
B. Pigeau, G. de Loubens, O. Klein, A. Riegler,	 F. Lochner, G. Schmidt	 \& L. W. Molenkamp, Nat. Phys. {\bf 7},  26  (2011).

\bibitem{hertel_gliga_2007}
R. Hertel, S. Gliga, M. F\"ahnle, and C.~M. Schneider, Phys. Rev. Lett. {\bf
  98},  117201  (2007).

\bibitem{lee_guslienko_2007}
K.-S. Lee, K.~Y. Guslienko, J.-Y. Lee, and S.-K. Kim, Phys. Rev. B {\bf 76},
  174410  (2007).

\bibitem{kravchuk07}
V.~P. Kravchuk, D.~D. Sheka, Y. Gaididei, and F.~G. Mertens, J. Appl. Phys.
  {\bf 102},  043908  (2007).

\bibitem{Berkov_Miltat_JMMM2007}
D. Berkov and J. Miltat, J. Magn. Magn. Mat. {\bf 320},  1238  (2008).

\bibitem{slonczewski96}
J.~C. Slonczewski, J. Magn. Magn. Mat. {\bf 159},  L1  (1996).

\bibitem{berger96}
L. Berger, Phys. Rev. B {\bf 54},  9353  (1996).

\bibitem{belavin75}
A.~A. Belavin and A.~M. Polyakov, JETP Lett. {\bf 22},  245  (1975).

\bibitem{gross78}
D.~J. Gross, Nucl. Phys. B {\bf 132},  439  (1978).

\bibitem{inpreparation}
S. Komineas and N. Papanicolaou, in preparation.

\bibitem{komineas98}
S. Komineas and N. Papanicolaou, Nonlinearity {\bf 11},  265  (1998).

\end{thebibliography}

\end{document}